\documentstyle[12pt,epsf]{article}

\newcommand{\be}{\begin{equation}}
\newcommand{\ee}{\end{equation}}
\newcommand{\beqa}{\begin{eqnarray}}
\newcommand{\eeqa}{\end{eqnarray}}
\newcommand{\bean}{\begin{eqnarray*}}
\newcommand{\eean}{\end{eqnarray*}}
\newcommand{\eqn}[1]{(\ref{#1})}
\newcommand{\nn}{\nonumber}

\newcommand{\del}{\partial}
\newcommand{\gapproxeq}{\lower .7ex\hbox{$\;\stackrel{\textstyle >}{\sim}\;$}}
\newcommand{\lapproxeq}{\lower .7ex\hbox{$\;\stackrel{\textstyle <}{\sim}\;$}}
\def\up#1{\leavevmode \raise.16ex\hbox{#1}}

\newcommand{\journal}[4]{{\sl #1 }{\bf #2} \up(19#3\up) #4}
\newcommand{\snabla}{\nabla\!\!\!\!/}
\def\sqr#1#2{{\vcenter{\vbox{\hrule height.#2pt
	\hbox{\vrule width.#2pt height#1pt \kern#1pt
			\vrule width.#2pt}
	\hrule height.#2pt}}}}
\def\square{\mathop\sqr68}
\newcounter{appendice}

\def\thebibliography#1{\section*{REFERENCES\markboth
	{REFERENCES}{REFERENCES}}\list
	{[\arabic{enumi}]}{\settowidth\labelwidth{[#1]}\leftmargin\labelwidth
	\advance\leftmargin\labelsep
	\usecounter{enumi}}
	\def\newblock{\hskip .11em plus .33em minus -.07em}
	\sloppy
	\sfcode`\.=1000\relax}

\begin{document}
\setlength{\unitlength}{1mm}
\begin{flushright}
\small DSF 52/96 \\
hep-th/9612168\\
\end{flushright}  

\vspace{1cm}
\begin{center}
{\Large\bf Three Dimensional Gross--Neveu Model on Curved Spaces}
\end{center}
\bigskip\bigskip\bigskip
\begin{center}
{{\bf Gennaro Miele} and {\bf Patrizia Vitale}}
\end{center}

\bigskip\bigskip
\begin{center}
{\it Dipartimento di Scienze Fisiche, Universit\`a di Napoli - 
Federico II -, and INFN\\ Sezione di Napoli, Mostra D'Oltremare Pad.  
20, 80125, Napoli, Italy}
\end{center}
\vspace*{2cm}
\begin{abstract}
The large $N$ limit of the 3-d Gross--Neveu model is here studied
on manifolds with positive and negative constant curvature. 
Using the $\zeta$--function regularization we analyze
the critical properties of this model on the spaces $S^2 \times S^1$
and $H^2\times S^1$. We evaluate the free energy density, the spontaneous 
magnetization and the correlation length at the ultraviolet fixed point. 
The limit $S^1\rightarrow R$, which is interpreted as the zero 
temperature limit, is also studied.

\end{abstract}
\vspace*{1cm}
\noindent
\begin{center}
{\it PACS number(s): 04.62.+v; 05.70.Jk; 11.10.Kk; 11.10.Wx}
\end{center}
\vspace*{2cm}
\noindent
e-mail: miele@na.infn.it; vitale@na.infn.it
\newpage
\baselineskip=.6cm

\section{Introduction}
\setcounter{equation}{0}
The Gross--Neveu model \cite{GN}, is an example of a fermionic
field theory which exhibits a non trivial ultraviolet (UV) fixed point
in 3 dimensions. From a field theoretic point of view such models, 
the non linear sigma--model is another one, provide at the fixed point, 
examples of
quantum field theories (QFT) which are {\it interacting}  and scale
invariant. In the Euclidean formalism such models describe classical
statistical systems undergoing second or higher order phase
transitions. Euclidean scale invariant QFT are then relevant from a
phenomenological point of view, since they describe experimentally
accessible phase transitions phenomena and the corresponding critical
exponents of such transitions have been calculated with great accuracy
in many realistic cases. With this respect, results have been obtained
mostly for theories which have a larger symmetry, conformal invariance
(particularly in 2 dimensions where the conformal algebra is infinite
dimensional). This is also the case for the nonlinear $\sigma$ model
in 3 dimensions, in the large N limit, as was shown in Refs.\cite{GRV,GV}.
We will see instead, that  the 3 dimensional GN model in the large N
limit is not conformally invariant at the fixed point. 
 
In this paper we study the  GN model, 
on 3-d manifolds of constant, non-zero curvature. The model exhibits
on $R^3$ a two--phase structure, the phase transition occurring for a
non--zero  value of the coupling constant (see for example \cite{zinn1} 
for a review). At the critical point there
is a symmetry breaking and the fermions become massive. The model
furnishes a description of phase transitions in classical
superconductors, as it was shown first in \cite{jacobs}, where the
stability conditions for the effective action in the large $N$ limit
are seen to imply the BCS gap equation. 

We consider the model on manifolds of the form ${\cal M}=\Sigma \times
S^1_{\beta}$ and analyze the limit $S^1_{\beta} {\rightarrow} R$ as
$\beta$ goes to infinity; $\Sigma$ is a 2--d manifold with constant,
non--zero, curvature. The interest for Euclidean field theories living
on manifolds of the form ${\cal M}=\Sigma \times S^1_{\beta}$ is due
to the fact that the radius of the circle $S^1_{\beta}$ can be
interpreted as the inverse temperature of some two--dimensional
statistical system. The action describing the model is then dependent
on the coupling constant present in the theory  and on the parameter
$\beta$. Phase transitions can occur with respect to both the
parameters. There is also a phenomenological reason to study critical
phenomena in curved spaces.  In fact, applying to a system an external
stress it is possible to deform its microscopic structure in such a
way to change the effective distance between points.  At a critical
point, universality suggests that the details of the microscopic
structures do not matter; but the system is still sensitive to the
deformation through the effective metric tensor density. Nevertheless,
the study of quantum effects for field theory at finite temperature in
curved space is an interesting subject by itself, since only in few
cases the quantum theory can be computed and in perspective one could
imagine some applications to cosmological scenarios. 

There exists a very reach literature on the 3--d GN model. The issue of 
critical 
exponents and $\beta$ function is addressed for example in 
Refs.\cite{grace1}--\cite{kivel}. In Refs.\cite{VKM}--\cite{krive} the 
effects of an external electromagnetic field are considered. The 
thermodynamical behaviour and the proof of $1/N$ renormalizability can be 
found in Refs.\cite{rosen1}--\cite{gat}. With respect to 
curvature induced symmetry restoration, see for example 
Ref.\cite{kiri} (GN model in 2 dimensions), and Ref.\cite{kane}.

The paper is organized as follows. In section 2 the properties of the
GN model are reviewed, whereas in section 3 the large $N$ limit
approximation is discussed. Section 4 is devoted to the flat space
analysis and in section 5 we introduce the $\zeta$-function
regularization. Thus in sections 6 and 7 the spaces $S^2_r \times S^1_{\beta}$
and $H^2_r \times S^1_{\beta}$ are analyzed at the critical coupling and the
limit $\beta \rightarrow 0$ is discussed. The asymptotic behaviour of
the correlation function for the above manifolds is studied in section
8. Finally in section 9 we give some concluding remarks. Some
technical details are left for the appendices. 

\section{The Gross--Neveu model in three dimensions}
\setcounter{equation}{0}
In this section we study the Gross--Neveu model on a Riemannian
manifold $({\cal M},g)$. The model is described in terms of a $O(n)$
symmetric action for a set of $N$ massless Dirac fermions. The
Euclidean partition function of the GN model in 3--dimensions in the
presence of a background metric $g_{\mu \nu}(x)$ is given by 
\be 
{\cal Z}[g]=\int {\cal D}[\psi]~ {\cal D}[{\bar\psi}]~ 
\exp\left\{-\int_{\cal M} d^3x~
\sqrt{g}~\left[{\bar \psi}_{i}(x) \snabla \psi_i(x) 
+ {q\over 2} ({\bar\psi}_{ i} \psi_i)^2\right]\right\}~~~,
\label{II.1}
\ee
where $i=1,2,\cdots,N$, $\snabla$ is the Dirac operator
on ${\cal M}$, and $q$ is the coupling constant\footnote{According to our
notation the Dirac matrices obey the following algebra: $\gamma_{\mu}
=  \gamma_{\mu}^{\dag}$, $\left\{ \gamma_{\mu}, \gamma_{\nu} \right\}
= 2 g_{\mu \nu}$, and $\mbox{Tr}\left(\gamma_{\mu}\right) = 0$.
Thus, the Dirac operator is antihermitian $\snabla^{\dag} = - 
\snabla$.}. 

The Dirac matrices on a generic 3-d manifold ${\cal M}$ are given in
terms of the Pauli matrices $\sigma_a$ by the expression 
\be
\gamma_\mu=V_{\mu,a} \sigma_a~~~,~~~~~\mbox{with}~~~\mu,a=1,2,3
\label{II.2}
\ee
where $V_{\mu,a}$ denote the {\it dreibein} defined by the equation
\be
g_{\mu\nu}=V_{\mu,a}(x) V_{\nu,b}(x) \delta_{ab}~~~. 
\label{II.3}
\ee
The {\it covariant derivative} $\nabla_{\mu}$ acting on a spinor field
is defined as \cite{zinn1,birrel} 
\be
\nabla_{\mu} = \partial_{\mu}
+ \Gamma_{\mu}(x)~~~,
\label{II.4}
\ee
where $\Gamma_{\mu}$ is the spin connection
\be
\Gamma_{\mu}(x) \equiv  { 1 \over 8} \left[ \sigma_{a},\sigma_{b}
\right] V^{\nu}_{a}\left( \nabla_{\mu} V_{\nu,b}\right)~~~.
\label{II.5}
\ee
The kinetic term of the action \eqn{II.1} is invariant under the
conformal transformation of the metric, $g_{\mu \nu}(x)\rightarrow
\Omega^{2}(x) g_{\mu \nu}(x)$. This can be easily checked observing
that the Dirac operator transforms according to \cite{birrel} 
\be
\snabla \rightarrow \left[\Omega^{-2}(x)\right] \snabla 
\left[\Omega(x)\right]~~~,
\label{II.6}
\ee
and assuming the spinors to be conformal densities of weight $-1$
\be
\psi_i(x)\rightarrow \Omega^{-1}(x) \psi_i(x)~~~.
\label{II.7}
\ee
On the contrary, the interacting term in \eqn{II.1} violates conformal 
invariance.

An equivalent way of rewriting the partition function \eqn{II.1}, but 
more suitable for our purposes is obtained by introducing an 
auxiliary scalar field $\sigma$, such that
\be
{\cal Z}[g]=\int {\cal D}[\psi]~{\cal D}[{\bar\psi}]~{\cal D}[\sigma] 
\exp\left\{-\int_{\cal M} d^3 x~\sqrt{g}~ 
\left[{\bar\psi}_{ i} (\snabla + \sigma) \psi_i 
-{1\over 2 q} \sigma^2\right]
~ \right\}~~~.
\label{II.8}
\ee
The new field has no effect on the dynamics of the theory, since from
the point of view of functional integration the integral over $\sigma$
merely multiplies the generating functional by an irrelevant constant.
We regularize the generating functional $\cal Z$ in the ultraviolet by
introducing a cut-off, $\Lambda$, in the momentum space. Before doing
that, let us compute the canonical dimensions of the fields and of the
coupling in our action. In mass units they result to be 
\be
[\psi]=[\bar\psi]= 1 \quad, \quad [\sigma]=1 \quad,
\quad \left[{1 \over q}\right]=1~~~.
\label{II.10}
\ee
By replacing the dimensional coupling constant $1/ q(\Lambda)$ with
the dimensionless ratio $\Lambda/ q(\Lambda)$, the regularized
partition function can be formally rewritten as 
\begin{eqnarray} 
{\cal Z}[g, \Lambda ]=\int {\cal D}_{\Lambda}[\psi]~
{\cal D}_{\Lambda} [{\bar\psi}]~
{\cal D}_{\Lambda}[\sigma]~
\exp\left\{-\int_{\cal M} d^3x~ \sqrt{g}~ 
\left[{\bar\psi}_{ i} (\snabla + \sigma) \psi_i -{\Lambda\over 2 q} \sigma^2 
\right]\right\},
\label{II.11}\nonumber\\
\end{eqnarray}
where ${\cal D}_{\Lambda}[\psi]=\prod\limits_{|k|<{\Lambda}} d\psi (k)$ and
similarly for the other fields.

\section{\bf The large $N$ limit}
\setcounter{equation}{0}
The GN model is not exactly solvable, so that we have to use some
approximation method. As for the non linear $\sigma$--model, the
existence of a non trivial UV fixed point shows that the large
momentum behaviour is not given by perturbation theory above 2
dimensions, where the theory is asymptotically free (see for
example Ref.\cite{zinn1}). Other techniques are required, like the $2+\epsilon$
expansion, which relies on the fact that the 2-d model is renormalizable 
in perturbation theory, or the $1/N$ expansion, which we will use in the 
paper. The model has been proven to be renormalizable in 3-d 
 in the $1/N$ expansion \cite{rosen1, parisi, gross}. 
	
In this limit, which means $N \rightarrow \infty$ keeping $N
{q(\Lambda)}$ fixed, the generating functional can be calculated using
the saddle point approximation. For this purpose we integrate over
$N-1$ fermion fields, rescale the remaining fields $\psi_{N}$,
${\bar\psi}_{N}$ to $\sqrt{N-1}~~ \psi_{N}$, $\sqrt{N-1}
~~{\bar\psi}_{N}$, respectively, and redefine $(N-1) {q(\Lambda)}$ as
${q(\Lambda)}$. Thus we get 
\begin{eqnarray} 
{\cal Z}[g,\Lambda,q(\Lambda)]=\int {\cal D}_{\Lambda}[\psi_{N}]~ 
{\cal D}_{\Lambda}[{\bar\psi}_{N}]~{\cal D}_{\Lambda}[\sigma]~
\exp\left\{ -(N-1)\mbox{Tr}\log_{\Lambda} (\snabla + \sigma)
\right\} \nonumber\\
\times \exp\left\{- (N-1)\int_{\cal M} d^3x~ \sqrt{g}~
\left[{\bar\psi}_N(\snabla +\sigma) \psi_N
-{\Lambda \over {2 q}}\sigma^2(x)\right]\right\}~~~.
\label{III.1}
\end{eqnarray}
In the limit $N \rightarrow \infty$ the dominating contribution to the
functional integral comes from the extremals of the action. For an
arbitrary metric $g_{\mu \nu}(x)$, these are obtained by extremizing
the action with respect to $\psi_{N}(x)$ keeping $\sigma(x)$ and
${\bar\psi}_N$ fixed and vice--versa. Hence, a set of equations
({\it gap equations}) is obtained 
\beqa
{\bar\psi}_{N} (\snabla -\sigma)=0~~~, \label{III.2}  \\
(\snabla +\sigma) \psi_N=0 ~~~,\label{III.3}  \\
{\bar\psi}_N \psi_N = {\Lambda \over {q(\Lambda)}} \sigma
- G_{\Lambda}(x, x; \sigma,g)~~~, 
\label{III.4}
\eeqa
where $G_{\Lambda}(x, x; \sigma,g)\equiv\langle x|(\snabla + \sigma)^{-1}|x
\rangle_{\Lambda}$ is the two-points correlation function of the
$\psi_N$-field, evaluated for $x \rightarrow x'$. Note that
the derivative in Eq. (\ref{III.2}) is acting on the left.
	
In the following analysis the generating functional 
${\cal Z}$ will be evaluated in the large $N$ limit, 
at the uniform saddle point
\be
\langle \sigma \rangle = m~~~,~~~
\langle \psi_N \rangle = b~~~,~~~
\langle {\bar\psi}_N \rangle={\bar b}~~.
\label{III.5}
\ee
The values $m$, $b$, ${\bar b}$ will be given by constant solutions
of gap equations. The quantities $b$ and ${\bar b}$ represent the
vacuum expectation value (v.e.v.) of fermion fields, while $m$, if
positive, can be regarded as the mass of the field fluctuations around
the vacuum. In the language of condensed matter physics $b$ is also 
addressed as {\it spontaneous magnetization}.

Once we find the saddle point solutions of the action, we can compute 
the generating functional of connected Green functions
${\cal W}=-\log({\cal Z})$ (the free energy), at the saddle point,
to the leading order in the $1 / N$ expansion
\be 
{{\cal W}[g, \Lambda,q(\Lambda)]}
=N \left[\mbox{Tr}\log_\Lambda(\snabla +m)-{\Lambda \over q}
\int_{\cal M} d^3x~ \sqrt{g}~{m^2} \right]~~~. 
\label{III.6}
\ee
The gap equations are the equations of motion of the classical
 field theory: the large $N$ limit of the Gross--Neveu model. The ground
state will then correspond to the solution which minimizes the free
energy. If the background  metric is homogenous, we expect the ground
state solution of the gap equations to be constant. 
	
\section{\bf Flat space $(R^3)$}
\setcounter{equation}{0}
Before solving the gap equations for curved spaces, it is worth noticing
that short distance divergences of the Green's function
$G_{\Lambda}(x, x; m^2,g)$ are independent of the curvature of the
space (the argument is completely analogous to the one given in
Ref.\cite{GRV} for scalar fields). This observation allows us to calculate
the critical value of the coupling constant in the simple case of flat
space  $R^3$, the critical coupling constant being, in fact, the value
of $q (\Lambda)$ which makes the divergences cancel in equation \eqn{III.4}. 

In the following we recall the two--phase structure of the GN model in 
flat space using Pauli--Villars regularization. Then we solve the gap 
equations for the physical parameters  $m$, $b$, and ${\bar b}$ and 
use these values (which must be independent of the regularization 
scheme) to determine the critical coupling constant in the $\zeta$--function 
regularization. 

The uniform saddle point is determined by the solution of the gap 
equations 
\beqa 
m b&=&0~~~,\label{IV.1}\\ 
{\bar b} b &=& {\Lambda \over q(\Lambda) } m 
- G_{\Lambda}(x, x; m,g)~~~,
\label{IV.2}
\eeqa
with 
\be 
G_\Lambda(x, x; m,g) = \lim_{x'\rightarrow x} 
\langle x| (\partial\!\!\!/ + m)^{-1} 
|x'\rangle = {{\Lambda m}\over {2\pi^2}} - {m^2 \over 4 \pi}~~~.
\label{IV.3}
\ee
Substituting the last expression into Eq. \eqn{IV.2} and 
recalling the definition of critical coupling constant we find
\be
{\Lambda\over q_c} = {\Lambda\over { 2 \pi^2}}~~~.
\label{IV.4}
\ee
Posing $M= (\Lambda/q) -(\Lambda/ q_{c})$,  Eq. \eqn{IV.2} becomes 
\be
\bar{b}b = Mm + {\Lambda\over q_{c}}m  - G_{\Lambda}~~~.
\label{IV.5}
\ee
Equation \eqn{IV.1} requires that either $m$ or $b$ be zero, or both. 
If $m$ is zero, \eqn{IV.2} implies $b=\bar b=0$. If $b$ is zero, we have
\be
Mm = -\frac{m^2}{4 \pi}~~~.
\label{IV.6}
\ee 
Hence, we can conclude that: $m$ and $b$ are zero at the critical point 
$M=0$; $m \ne 0,~b=0$ for $M<0$, while $m=b=0$ when $M>0$. The point $M=0$ 
is where the phase transition takes place, while for $q> q_{c}$ the 
fermions acquire mass and the symmetry is spontaneously broken.
In the next section we use $m_c=b_c=0$ to determine the critical 
coupling constant in the $\zeta$--function regularization.
	
\section{$\zeta$--function regularization}
\setcounter{equation}{0}
We will use throughout the paper the $\zeta$--function regularization. In 
fact, it results to be more tractable when we are on curved spaces, even 
though other regularizations such as the Pauli--Villars have
a more immediate physical meaning. 
Since the critical value of the
coupling constant at which the theory becomes finite is independent of
the background metric, we compute this critical coupling on $R^3$. 

Let us consider the squared Dirac operator in $d$ dimensions
\be
\Delta_{1/2} \equiv
\snabla^{\dag} \snabla= - \nabla^2_{1/2} 
+ {{\cal R}\over 4} ~~~.
\label{V.1}
\ee
Note that $\Delta_{1/2}$ is not the conformal spin 1/2
Laplacian, which is instead the combination \cite{parker}
\be
\square_{1/2}= \Delta_{1/2} - {1 \over 4(d-1)} {\cal R}~~~,\label{V.2}
\ee
where ${\cal R}$ denotes the Ricci scalar. 

Given the eigenvalues, $\lambda^2_{n}+m^2$, of the operator $(\Delta_{1/2}
+ m^2)$ and an orthonormal basis of corresponding eigenvectors 
$\left\{ \psi_{n}(x)\right\}$, the local $\zeta$--function
is defined as 
\be
\zeta(s,x)=\sum_{n} (\lambda^2_n+m^2)^{-s} |\psi_n(x)|^2~~~,
\label{V.3}
\ee
where the sum includes degeneracy, and in case $m=0$, $\lambda_{n}$ 
must be non vanishing. If the eigenvalues are continuous,
the sum is replaced by an integral. The Green's function for the
operator $\snabla +m$ can be defined through the $\zeta$--function
\eqn{V.3}. 
Observing that
\begin{eqnarray}
\langle x| (\snabla + m)^{-1} |x\rangle &=& \sum_{n,l} \langle x|\psi_n\rangle 
\langle \psi_n| (\snabla + m)^{-1} |\psi_l\rangle 
\langle \psi_l|x\rangle \nn\\
&=& \sum_n |\psi_n(x)|^2 \frac{(-i \lambda_n + m)}{\lambda_n^2 
+m^2}
\label{V.4}
\end{eqnarray} 
and recalling that the eigenvalues of the Dirac operator $\snabla$ always 
appear in pairs $\pm i \lambda_n$, we regularize the Green's function 
of the operator $\snabla + m$ as
\be
G_s(x,x;m,g)= m\langle x|(\Delta_{1/2}+m^2)^{-s}|x \rangle=m\zeta(s,x)~~~,
\label{V.5}
\ee
and
\be 
G(x,x;m,g)=m~\lim_{s \rightarrow 1} \zeta(s,x)~~~.
\label{V.6}
\ee
On homogeneous spaces such as the ones we will be considering in this paper,
$\zeta(s,x)$ turns out to be independent of $x$.

On $R^3$ the gap equation \eqn{III.4} becomes, in this regularization
scheme,
\beqa
m~ \lim_{s\rightarrow 1}{1 \over {q (s)}} & = & {\bar b} b 
+ m\lim_{s\rightarrow 1} \zeta(s,x)
~=~ {\bar b} b + m\lim_{s\rightarrow 1} 
\int {d^3k \over (2 \pi)^3} {1\over (k^2+m^2)^s}\nonumber\\
&=&{\bar b} b + {m\over 2 \pi^2}
\lim_{s\rightarrow 1} \int_0^\infty dt~
{t^{s-1} \over \Gamma(s)} {\int k^2~ dk~e^{-(k^2+m^2)t}}~~~, 
\label{V.7}
\eeqa
where the regularized coupling $\Lambda /q(\Lambda)$ in the
Pauli-Villars regularization has been replaced by
$1/q(s)$ in the $\zeta$--function regularization.
Here we have used the Mellin transform to analytically continue the
$\zeta$--function. Note that
\be
\zeta (s)= \int {d^3k \over (2 \pi)^3 (k^2+m^2)^s}~~~,
\label{V.8}
\ee
has no pole at $s=1$. It is now easy to verify that
\be
\lim_{s\rightarrow 1} {m \over q (s)}
={\bar b} b - \lim_{s\rightarrow 1}
{m^{-2s+4} \over {(4 \pi)}^{ 3 \over 2}}
{\Gamma(s-{3 \over 2})\over {\Gamma (s)}}=
{\bar b} b+{m^2 \over 4 \pi}~~~. 
\label{V.9}
\ee
Using the critical values of $b$, ${\bar b}$ and $m$, found 
in the previous section (being physical quantities they are regularization 
scheme independent), we get, in  the $\zeta$--function regularization
\be 
{1 \over {q_c}}=0~~~.
\label{V.10}
\ee
Hereafter we use this value of the critical coupling, since 
it is independent of the background metric.
	
At the critical point the large $N$ limit of the free energy 
\eqn{III.6} becomes then 
\be 
{\cal W}_c(g)= {N \over 2} \log\det(\snabla +m_c)~~~.
\label{V.11}
\ee
Observing that $\log\det(\snabla +m_c) = (1/2) \log\det(\Delta_{1/2} +
m^2_c)$ one can define the free energy density $w_c(g)$ through the
$\zeta$--function \eqn{V.3}. We have 
\be
\det(\Delta_{1/2}+m^2_c)=\cases {0, &if 
$\mbox{dim}(\mbox{ker}[\Delta_{1/2}+m^2_c])\not=0$\cr
e^{-\zeta^{\prime}(0)}, &if $\mbox{dim}(\mbox{ker}
[\Delta_{1/2}+m^2_c])=0$\cr}~~~.
\label{V.12}
\ee
The free energy density at the critical point is then regularized as 
\be
w_c(g) = \cases {0, &if $\mbox{dim}(\mbox{ker}[\Delta_{1/2}+m^2_c])\not=0$\cr
-{1 \over 2}\zeta^{\prime}(0), &if $\mbox{dim}(\mbox{ker}[
\Delta_{1/2}+m^2_c])=0$\cr}~~~.
\label{V.13}
\ee
This quantity is indeed a density, since the $\zeta$--function
defined in Eq.\eqn{V.3} contains an inverse volume factor through the
squared modulus of eigenvectors. 

We wish to stress at this point a major difference with the non--linear 
sigma model. It was found in Ref.\cite{GRV} that the free energy for the non 
linear $\sigma$ model at the 
critical point is equal to 
\be
{\cal W}_{c}= {N\over 2}\log\det(\square+m^2_c)~~~,
\label{V.14}
\ee
where $\square$ is the conformal scalar Laplacian \cite{birrel}.
Using this fact (of the Laplacian being conformal) it was shown that
the free energy is conformally invariant. At a first sight equation
\eqn{V.14} looks identical to the expression \eqn{V.13}. But, we have
seen that $\Delta_{1/2}$ is not the conformal Laplacian, then the
arguments used in Ref.\cite{GRV} do not apply and we can conclude that
the free energy of the Gross--Neveu model at the critical point is
scale invariant but not conformally invariant. 

\section{The Manifold $S^2_r \times S^1_\beta$}
\setcounter{equation}{0}
In this section we study the large $N$ limit of the Gross--Neveu model
on the manifold $S_r^2 \times S_\beta^1$ ($r$ and $\beta$ are the two
radii). We parametrize this space by $x^{\mu} \equiv (\tau,~
\chi,~\theta)$, 
where $0\le\tau< 2\pi$, $-\pi/2\le\chi\le \pi/2 $,
and  $0\le\theta< 2\pi$. The metric tensor is then defined as 
\be
g_{\mu \nu} dx^\mu \otimes dx^\nu=r^2~\cos^2\chi~d\tau \otimes d\tau
+ r^2 d\chi \otimes d\chi+ \beta^2 d\theta \otimes d\theta~~~. 
\label{VI.1}
\ee
Using the definition \eqn{II.2}, the Dirac matrices on $S^2_r\times 
S^1_{\beta}$ are given in terms of the Pauli matrices $\sigma_a$,  by  
\be
\gamma_{1} = r \cos\chi ~ \sigma_1~~~,~~~~~~~~~
\gamma_{2} = r \sigma_2~~~,~~~~~~~~\gamma_3=\\ \beta \sigma_3
~~~.
\label{VI.2}
\ee
The spin connection \eqn{II.5} results to be 
\be
\Gamma_{\mu}(x) =
- {i \over 2} \sigma_{3} \sin{\chi}~\delta_{1\mu}~~~,
\label{VI.3}
\ee
while the  covariant derivative 
$\nabla_{\mu}$ acting on a spinor field is given by \eqn{II.4}. 
Substituting \eqn{VI.2} and \eqn{VI.3} in the Dirac equation 
\be
\snabla \psi_{\lambda} = i \lambda \psi_{\lambda}~~~.
\label{VI.4}
\ee
we find the eigenvalues of $\snabla$ to be $i 
\lambda_{l,n}^{\pm}$ with
\be
\lambda_{ln}^{\pm}=\pm \sqrt{
\frac{(2n+1)^2}{4\beta^2} + {(l+1)^2 \over r^2 }}~~~~~l=0,1,..~~~n=0, \pm 1,..
\label{VI.5}
\ee
and degeneracy $2(l+1)$; note that there are no zero modes. The
details of calculation are given in Appendix A. 

The eigenvalues of $\Delta_{1/2} + m^2$ are then given by
$|\lambda_{ln}|^2 + m^2$, with degeneracy $4(l+1)$. The
$\zeta$--function \eqn{V.3} for this operator is 
\be
\zeta(s,m) = {1 \over 4 \pi^2 \beta r^2} 
\sum_{n=-\infty}^{\infty} \sum_{l=0}^{\infty} 
\left[ {(2 n+1)^2 \over 4 \beta^2} + {(l+1)^2 \over r^2} +m^2\right]^{-s}
(l+1)~~~,
\label{VI.6}
\ee
while the gap equations \eqn{III.3} and \eqn{III.4} become 
\beqa
(\gamma^{\mu} \Gamma_{\mu} + m) b&=&0~~~,\label{VI.7}\\
{\bar b} b &=& m \lim_{s\rightarrow 1}\biggl\{ \frac{1}{q(s)} 
-\zeta(s,m)\biggr\}~~~.
\label{VI.8}
\eeqa
The first one yields $b=0$, so that we have to solve 
\be
0 = m \lim_{s\rightarrow 1}\biggl\{ \frac{1}{q(s)} -\zeta(s,m) 
\biggr\}~~~.
\label{VI.9}
\ee
Since $\lim_{s\rightarrow 1} 1/ q(s) = 0 $ and $ \lim_{s\rightarrow 1 }
\zeta (s, m=0)  $ is well defined, $m=0$ is a solution. For $m\ne 0$
there is no value of $m$ such that $\zeta(s,m) = 0$. Then $m_c=0$, $b_c=
{\bar b}_c=0$ are the critical values of the physical mass and the vacuum
expectation value of the $\psi$, $\bar\psi$ fields on $S^2_r \times
S^1_\beta$, in the large N limit. Note that $m_c=0$ is not in
contradiction with the fact that the correlation length has to be finite
due to the {\it finite size} of the background. In fact we have already
observed that the operator $ \Delta_{1/2}$ has no zero modes, so that the
smallest eigenvalue of $\Delta_{1/2} + m_c^2$ is non zero. 

To evaluate the large N limit of the free energy density at
criticality, we have to take the derivative of the $\zeta$--function at
zero and substitute the value of $m_c$ that we have just found,
in Eq. \eqn{V.13}: 
\be
w_c= {-1\over 8 \pi^2 r^2 \beta}
\lim_{s\rightarrow 0} \Biggl\{\frac{d}{ds} 
\sum_{n=-\infty}^{\infty} \sum_{l=0}^{\infty} 
\left[ {(2 n+1)^2 \over 4 \beta^2} + {(l+1)^2 \over r^2} +m_c^2\right]^{-s}
(l+1)\Biggr\}. 
\label{VI.10}
\ee 
We first take the Mellin transform of $\zeta(s,m_c)$
\be
\zeta (s,m_c)= {\beta^{-2s-1}\over 4 \pi^2 r^2
\Gamma(s)} \int^{\infty}_0 dt~ t^{s-1}
\sum_{n=-\infty}^{\infty} \sum_{l=1}^{\infty} l \exp \Biggl\{-
\left[ \left( n+{1 \over 2}\right)^2  + l^2{\beta^2 \over r^2} +m_c^2\beta^2 
\right]t\Biggr\};
\label{VI.11}
\ee
then use a specialization of the Poisson sum formula for the sum over 
$l$ (the formula is obtained in Appendix B), which allows us to exchange the 
sum over $l$ with the integral in $t$
\be
\sum_{l=1}^{\infty} l \exp \left\{-l^2{\beta^2 t\over r^2} \right\}=
{r^3\over 2{\sqrt 
\pi} \beta^3 } t^{-3/2} \int_{-\infty}^{\infty} dx ~ x~ \cot(x)~ 
\exp\left\{-x^2{r^2\over\beta^2 t}\right\} \label{VI.12}~~~,
\ee 
and we get
\beqa
\zeta (s,m_c)= {r \over 8 \pi^{5/2} \Gamma(s)} \beta^{-2s-4} 
\int^{\infty}_0 dt~ t^{s-5/2}
\sum_{n=-\infty}^{\infty} P\int_{-\infty}^{\infty} dx ~\left[x\cot(x) -1
\right]\nonumber\\
\times\exp\left\{-\left[ \left(\left(n+{1\over 2}\right)^2 
+ m_c^2 \beta^2 \right)t + x^2{r^2 \over \beta^2 t} \right]
\right\} \nonumber\\
+ {\beta^{-2s-3}\over 8 \pi^2 \Gamma(s) } 
\int^{\infty}_0 dt~ t^{s-2}
\sum_{n=-\infty}^{\infty}  
\exp\left\{-\left[ \left(n+{1\over 2}\right)^2 
+ m_c^2 \beta^2 \right]t\right\} \equiv A+B.
\label{VI.13}
\eeqa
Here we extracted the part of the integral in $x$ which would become 
divergent once we exchange the order of integration. We find also
convenient not to put $m_c=0$ until the end of the calculation.

To evaluate the contribution of $A$ to the derivative of the $\zeta$ 
function we first perform the integral in $t$. We get
$$
\lim_{s\rightarrow 0} \left[{d\over ds} A\right]= 
{1 \over 4 \pi^2 r^2 \beta}
\sum_{n=0}^{\infty} P\int_0^{\infty} {dx \over x^3}~\left[x\cot(x) -1 
\right]
$$
\be
\times \left\{1+2x {r\over\beta} 
\left[\left(n+{1\over 2}\right)^2 +m_c^2\beta^2\right]^{1/2}\right\}
\exp\left\{- 2x {r\over\beta} \left[
\left(n+{1\over 2}\right)^2 +m_c^2\beta^2\right]^{1/2}
\right\}.
\label{VI.14}
\ee
The principal value of the integral can be evaluated using the method 
of residua, since it has only simple poles. Thus, for 
$m_c=0$, we find
\be
\lim_{s\rightarrow 0} \left[{d\over ds} A\right]_{m_c=0} 
=  {3\zeta_{R}(3)\over 32 \pi^4 \beta^3} + {1 \over 8 \pi^2 r^2 \beta}
\sum_{n=1}^{\infty} {(-1)^n \over n} \mbox{cosech}^2 
\left(n\pi{\beta\over r}\right)~~~,
\label{VI.15}
\ee
where $\zeta_R (z)$ denotes the Riemann $\zeta$--function.

The second integral in \eqn{VI.13} is easily computed by using the Poisson 
formula for the sum over $n$ \cite{whitt}
\be
\sum_{n=-\infty}^{\infty} \exp \left\{ - \left(n+{1\over 2}\right)^2  
t\right\}=
{\sqrt{\pi\over t}} \left[ 1 + 2 \sum_{n=1}^{\infty} (-1)^n \exp\left\{-{\pi^2 
n^2\over t}\right\} \right]~~~.
\label{VI.16}
\ee
Thus one finds
\beqa
B&=& {1\over 4 \pi^{3/2} \Gamma(s) }  \beta^{-2s-3}
\sum_{n=1}^{\infty}(-1)^n
\int^{\infty}_0 dt~ t^{s-5/2}
\exp\left\{ -\left[{\pi^2 n^2\over t}  m_c^2 \beta^2 t  
\right]\right\} \nn\\  
&+& 
{1\over 8 \pi^{3/2} \Gamma(s) }  \beta^{-2s} m_c^{3-2s} \Gamma\left(
s-{3 \over 2}\right)~~~.\label{VI.17}
\eeqa
In the limit $s \rightarrow 0$ 
$m_c=0$, we find then
\be
\lim_{s\rightarrow 0} \left[{d\over ds} B\right]_{m_c=0}
= {1\over 8\pi^4 \beta^3} \sum_{n=1}^{\infty} {(-1)^n\over n^3}= - 
{3\zeta_{R}(3)\over 32 \pi^4 \beta^3} ~~~.
\label{VI.18}
\ee
Summing \eqn{VI.15} with \eqn{VI.18} we finally get  
the regularized free energy density at the critical point 
\be
w_c= -{ 1 \over 16 \pi^2 r^3 } \left({r \over \beta}\right)
\sum_{n=1}^{\infty} {(-1)^n \over n}~ \mbox{cosech}^2 
\left(n\pi{\beta\over r}\right)~~~.
\label{VI.19}
\ee
It can be checked that the series is convergent to a negative value for
any finite value of $\beta/ r$, so that the free energy density is
positive definite.  The series can be evaluated numerically. In Figure 
1 we plot the free energy density of Eq.\eqn{VI.19} as a function of the ratio 
$\beta /r$ (up to the overall
factor $1/16 \pi^2 r^3$) 

The limit 
$\beta\rightarrow \infty$ , which corresponds to zero temperature, 
yields $w_c= 0$, in agreement with the 
result that we would get by direct calculation on $S^2_r \times R$.
The limit $r\rightarrow \infty$, which corresponds to the manifold 
$R^2\times S^1_{\beta}$, yields
\be
\lim_{r\rightarrow \infty} w_c= -{1\over 16 \pi^4\beta^3} \sum_{n=1}^{\infty} 
{(-1)^n\over n^3}= {3\zeta_{R}(3)\over 64 \pi^4 \beta^3} ~~~, \label{VI.20}
\ee
where $\zeta_{R}(3)=1.20205...$. This represents the free energy density 
for the GN model on flat space, 
at finite temperature $1/\beta$. As we can see it goes to zero with 
temperature. To our knowledge this result was first obtained in 
\cite{VKM}, where the free energy density of the GN model is evaluated on 
$R^2\times S^1$ in the presence of a magnetic field. The limit of zero 
magnetic field is in agreement with \eqn{VI.20} once we observe that 
$m_c$ is zero in this limit.

Finally, we wish to note that $\zeta(0,m_c)=0$ in agreement with the
scale invariance of the model at criticality. 

\section{The Manifold $H^2_r \times S^1_\beta$}
\setcounter{equation}{0}
We consider the product manifold $H^2_r \times S^1_\beta$. $H^2_r$ is a 
2--dimensional {\it pseudosphere}, namely, it is obtained 
globally embedding  a hyperboloid in $R^3$ endowed with
Minkowskian metric. Either sheet of the hyperboloid models an infinite 
space--like surface (hence with Riemannian metric) without boundary. 
This surface has constant negative curvature and it is the only simply 
connected manifold with this property \cite{voros}.
We parametrize $H^2_r$ as
\be
H^2_r=\{ z = (x, y),~~ x\in R,~~ 0 <y<\infty \}~~~,
\label{VII.1}
\ee
while the circle $S^1$ of radius $\beta$ is parametrized as before
by $\theta$, $0 \leq \theta <2\pi$. The scalar curvature of $H^2_r$
is ${\cal R} = -2/r^2$, where $r$ is a constant positive parameter.
The metric tensor on the whole manifold is then given by
\be
g_{\mu \nu} dx^\mu \otimes dx^\nu
={r^2 \over y^2} \biggl(dx\otimes dx + dy \otimes dy
\biggr) + \beta^2 d\theta \otimes d\theta~~~, 
\label{VII.2}
\ee
where $x^{\mu}\equiv(x,y,\theta)$ with $\mu=1,2,3$.
The Dirac matrices on $H^2_r \times S^1_\beta$ are given in terms of flat 
Dirac matrices   by
\be
\gamma_1=  {r \over y} \sigma_1~,~~
\gamma_2=  {r \over y} \sigma_2~,~~
\gamma_3=  \beta \sigma_3 ~,
\label{VII.3}
\ee
while the spin connection defined in \eqn{II.5}, is
\be
\Gamma_{\mu}= {i\over 2y} \sigma_3 \delta_{1 \mu}~~~.
\label{VII.4}
\ee
To find the spectrum of the Laplacian, we could  in principle proceed
as in the previous case, by using the algebraic method described in
Appendix A. However, in order
to write the $\zeta$--function all that we need is the heat
kernel of the Laplacian. Thus, we use the algebraic method only to
establish that the eigenvectors of the Dirac operator (and hence of
its square) are of the form 
\be
|\psi_{n,k}\rangle=  |\chi_k\rangle \otimes |\phi_n\rangle~~~,
\label{VII.5}
\ee
where 
\be
|\phi_n\rangle= {1 \over \sqrt{2 \pi \beta}}
\exp \left\{i \left( n+{1 \over 2}\right) \theta\right\}
\label{VII.6}
\ee
are scalars depending on $\theta$ only, while $|\chi_k\rangle$
are spinors depending on $x$, $y$. They are more
explicitly described in Appendix A. This means that the 
Laplacian can be factorized as
\be
\Delta_{1/2}(H^2_r \times S^1_\beta)= \Delta_{1/2}(H^2_r)+\Delta_0 
(S^1_\beta)
\label{VII.7}
\ee
and the heat kernel is just the product of the heat 
kernels of the Laplacians on $H^2_r$ and $S^1_\beta$, respectively,
\be
h(t; x, y, \theta)= h_{H^2_r}(t; x, y)  h_{S^1_\beta}(t;\theta)~~~.
\label{VII.8}
\ee
The heat kernel for the spin 1/2 Laplacian on $H^2_r$ is \cite{doker}
\begin{eqnarray}
h_{H^2_r}(t; z, z')=
{2r(4\pi t )^{-3/2} \over \sqrt{\cosh (d/r) 
+1}} \int_{d/r}^\infty dw~ {w~\cosh (w/2) \over \sqrt{\cosh(w) 
- \cosh (d/r)}}      
\exp\left\{ {-w^2r^2\over 4t}\right\}
\label{VII.9}
\end{eqnarray}
where $d$ is the geodesic distance among points on $H^2_r$, 
\be
\cosh\left({d \over r}\right)=  1 + {|z-z'|^2 \over 2y y'}~~~,       
\label{VII.10}
\ee
and $z=x+iy$. In the limit $d\rightarrow 0$ Eq.\eqn{VII.9}  becomes
\be
h_{H^2_r}(t; z= z')=\frac{r}{2(\pi t )^{3/2}} 
	\int_0^\infty dw~w~\coth(w)       
\exp\left\{ {-w^2r^2\over t}\right\}~~~, \label{VII.11}
\ee
The equal points heat kernel of the scalar Laplacian on $S^1_\beta$ 
(the spectrum is found in Appendix A)
is
\be
h_{S^1_\beta}(t;\theta= \theta')= {1 \over 2 \pi \beta}
\sum_{n=-\infty}^{\infty} \exp\left\{-\left(n+{1\over 2} \right)^2
{t \over \beta^2} \right\}~~~. \label{VII.12}
\ee
We are now ready to write the spectral $\zeta$--function for the
operator $\Delta_{1/2} + m^2$ on the product manifold. It is defined
in terms of the heat kernel as 
\be
\zeta(s,m)={1\over \Gamma(s)} \int_0^\infty dt~t^{s-1} h_{H^2_r}(t; z=z') 
h_{S^1_\beta}(t;\theta=\theta')~\exp\left\{-m^2 t\right\}~~~. \label{VII.13}
\ee
On $H^2_r \times S^1_\beta$
the gap equations \eqn{III.3} and \eqn{III.4} become 
\beqa
(\gamma^\mu \Gamma_\mu + m) b&=&0~~~,\label{VII.14}\\
{\bar b} b &=& m \lim_{s\rightarrow 1}\biggl\{ \frac{1}{q(s)} 
-\zeta(s,m) \biggr\}~~~.\label{VII.15}
\eeqa
The first one yields $b=0$, so that we have to solve 
\be
0 = m \lim_{s\rightarrow 1}\biggl\{ \frac{1}{q(s)} -\zeta(s,m) 
\biggr\}~~~.                                
\label{VII.16}
\ee  
By rescaling
$t \rightarrow t\beta^2$, the $\zeta$-function \eqn{VII.13}
becomes
\beqa
\zeta(s,m)&=& { r\over 4\pi^{5/2} \Gamma(s)} \beta^{2s-4}
\int_0^\infty dt~t^{s-5/2}  
\Biggl\{ 
\int_0^\infty dw ~w~\coth(w)   \nn\\
&\times& \sum_{n=-\infty}^{\infty}
\exp \left[ -  {w^2r^2\over \beta^2 t} - 
\left( \left(n+{1\over 2}\right)^2 + m^2 \beta^2\right)t \right]\Biggr\}~.  
\label{VII.17}
\eeqa
One can check that, since $\lim_{s\rightarrow 0} \zeta(s, m=0)$ is 
finite, $m=0$ is a solution of \eqn{VII.16}. Thus, at the critical point
we have 
\be
m_c=b_c=0~~~. \label{VII.18}
\ee
To find the free energy density at criticality we have to consider
the derivative of the $\zeta$--function at $s=0$, and then evaluate it 
for the critical value of $m$ just found.
The calculations go along the same lines as those for $S^2_r\times S^1_\beta$:
each time we exchange integrals and series we have to verify 
that no divergences are introduced, and in that case
eventually regularize them.

We have 
\beqa
\zeta(s,m_c)&=&{r \over 4\pi^{5/2} \Gamma(s)} \beta^{2s-4}
\left\{\int_0^\infty  dt~ t^{s-5/2}   
\sum_{n=-\infty}^{\infty} 
\int_0^\infty dw \quad \left[w~\coth(w) -1 \right]\right.\nn\\ 
&\times& \exp\left\{-{w^2 r^2 \over \beta^2 t} 
-\left[\left(n+{1\over 2}\right)^2 
+ m_c^2 \beta^2 \right]t\right\}\nn\\    
&+&\left.
\int_0^\infty  dt~ t^{s-5/2} \sum_{-\infty}^{\infty}   
\int_0^\infty dw~ 
\exp \left\{-{w^2 r^2\over \beta^2 t} -\left[\left(n+{1\over 2}\right)^2 
+ m_c^2 \beta^2 \right]t\right\}\right\}\nn\\   
&\equiv&  A~+B~~~.\label{VII.19}   
\eeqa
The derivative of the first integral in \eqn{VII.19},
evaluated at $s\rightarrow 0, m_c=0$, gives 
\be
\lim_{s\rightarrow 0}\left[{d\over ds} A\right]_{m_c=0}=
{1\over 8 \pi^2 r^3 } \int_0^\infty {dw \over w^2} \left[\coth(w) 
-{1\over w}\right] \left[ 1- { wr\over\beta} \coth\left({wr\over\beta}  
\right) \right] {r/\beta \over \sinh{(w~r/ \beta)}}
\label{VII.20}
\ee
The derivative of the second integral in \eqn{VII.19}, at $s\rightarrow 0$, 
$m_c=0$,
gives
\be
\lim_{s\rightarrow 0}\left[{d\over ds} B\right]_{m_c=0}= 
{1\over 8 \pi^4 \beta^3} \sum_{n=1}^\infty {(-1)^n \over n^3}~~~.
\label{VII.21}
\ee
Putting together \eqn{VII.20} and \eqn{VII.21} we get the expression
of the free energy density at the critical point to be 
\beqa
w_c&=&-{1\over 16\pi^2 r^3 } \left\{
 \int_0^\infty {dw \over w^2} \left[\coth(w) 
-{1\over w}\right] \left[ 1- { wr\over\beta} \coth\left({ wr\over\beta}  
\right)\right] {r/\beta\over \sinh{(w~r/\beta)} } \right.\nn\\
&- & \left.{3\zeta_R(3)\over 4 \pi^2}\left({r \over \beta}\right)^3
\right\} ~~~.
\label{VII.22}
\eeqa
The integral can be evaluated numerically and the free energy density $w_c$
results to be positive definite. In Figure 2 
the free energy density of \eqn{VII.22} 
is plotted as a function of the ratio $\beta/r$, up to the overall 
factor $1/16 \pi^2 r^3$).

As for the manifold $S^2_r\times S^1_\beta$, the limit of zero temperature 
can be evaluated analytically and we get
\be
\lim_{\beta\rightarrow \infty} w_c=0~~~.
\ee 
The limit $r\rightarrow \infty$, which corresponds to the manifold 
$R^2\times S^1$ yields
\be
\lim_{r\rightarrow \infty} w_c ={3\zeta_R(3)\over 64 \pi^4 \beta^3} ~~~.
\label{VII.23}
\ee 
As expected, we get the same result that we found in \eqn{VI.20} 
for the limit $S^2\rightarrow R^2$.

\section{Asymptotic behaviour of the correlation function}
\setcounter{equation}{0}
To understand the effects of the curvature on the second order phase
transition exhibited from the model on flat space, we investigate the
behaviour of the two--points 
critical correlation function as the distance among
points goes to infinity. The correlation length which characterizes
such a behaviour, can also be regarded as the inverse of the smallest
eigenvalue of the operator under consideration. For this reason it is
then equivalent, but easier, to study the correlation function of the
squared Dirac operator instead than the Dirac operator itself. 

When a second order phase transition occurs, the correlation length
diverges and the two point Green's function follows a power law for
large distances. This is what happens for the GN model on $R^3$ at the
critical point. We don't analyze the case of $S^2_r\times S^1_\beta$, 
because finite size of the manifold in all directions forces the 
correlation length to be finite.
 On the other hand the manifold $H^2_r\times S^1_\beta$ has two
non compact directions; it is then meaningful to study the asymptotic
behaviour of the Green's function at criticality, when the distance
among points on $H^2$ diverges. We will also study such a behaviour
in the limit of zero temperature (no compact directions at all). 

The 2--point Green's function of the operator $\Delta_{1/2} + m^2$ is
given by 
\be
G(z,\theta, z',\theta')= \int_0^\infty dt~ 
h_{H^2_r}(t;z,z')h_{S^1_\beta} 
(t;\theta,\theta') \exp\left\{-m^2 t\right\}~~~.
\label{VIII.1}
\ee
Since we are interested in the asymptotic behaviour of the Green's 
function in the $H^2$ direction we can fix $\theta=\theta'$. 
Moreover, at criticality it is $m=0$.  Using \eqn{VII.9} and \eqn{VII.12}
we have then
\beqa
G(z, z',\theta)&=& {r \beta^{-1}\over (2 \pi)^{5/2} } 
{1\over \sqrt{\cosh (d/r) 
+1}} \int_0^\infty dt ~
t ^{-3/2}  \int_{d/r} ^\infty dw {w~\cosh (w/2) \over \sqrt{\cosh(w) - 
\cosh (d/r)}} \nn\\      
&\times& \sum_{n=0}^\infty \exp\left\{ -{w^2r^2\over 4t} -\left(n+{1\over 2}
\right)^2 {t\over \beta^2} \right\}~~~.
\label{VIII.2}       
\eeqa
After performing the integral in $t$ and the sum over $n$ we get
\beqa
G(z, z',\theta)= {\beta \over 4 \sqrt{2} \pi^{2} } {1\over \sqrt{\cosh (d/r) 
+1}} 
\int_{d/r} ^\infty dw {w~\cosh (w/2) \mbox{cosech} (wr/2\beta)\over 
\sqrt{\cosh(w) - 
\cosh (d/r)}}
\label{VIII.3}
\eeqa
In the limit  $d \rightarrow \infty$, with $r$ and $\beta$ finite, we can 
approximate the integral as 
\be
G(z, z',\theta)\sim {\sqrt{2} \over 4 \pi^{2} \beta} 
\exp\left\{-{d\over 2r}\right\}  
\int_{d/r}^\infty dw~ \left[w -{d \over r}\right]^{-1/2} 
\exp\left\{-{wr\over 2\beta}\right\}~~~.
\label{VIII.4}
\ee
We finally get
\be
G_{H^2_r \times S^1_{\beta}}
(z, z',\theta) \sim {1 \over 4 \pi^{3/2} \sqrt{r \beta}} 
\exp\left\{-{d \over 2}\left({1\over r}+{1\over \beta}\right)\right\}~~~. 
\label{VIII.5}
\ee
As one can see, the correlation length at criticality is finite
\be
\xi=2 \left({1\over r}+{1\over \beta}\right)^{-1}~~~,
\label{VIII.6}
\ee 
namely, it is proportional to the two scales of the theory, $\beta$ and 
$r$.

To analyze the zero temperature case one cannot use the results 
\eqn{VIII.5},\eqn{VIII.6}, which were obtained assuming $\beta$ 
finite, but rather Eq.\eqn{VIII.3} in the limit 
$\beta\rightarrow \infty$. Thus we get
\be
G(z, z',x_o)={\sqrt{2}\over r\pi^2} {1\over \sqrt{\cosh(d/r) +1}} 
\int_{d/r}^\infty {dw \over w}
{\cosh(w/2) \over \sqrt{\cosh(w) -\cosh(d/r)}}~~~,
\label{VIII.7}
\ee
where $x_o=x_o'$ is the coordinate on $R$. We now take the limit 
$d\rightarrow \infty$, $r$ finite, and we get
\be
G_{H^2\times R}(z, z',x_o)\sim {2\over \pi^2\sqrt{rd}} \exp\left\{-{d\over 
2r}\right\}~~~.
\label{VIII.8}
\ee 
As we can see, even in the case of a manifold which is non compact in all 
directions, we get for the asymptotic Green's function an exponential 
behaviour, and hence a finite correlation length, which is proportional to 
the radius of curvature $r$.

\section{Conclusions}
\setcounter{equation}{0}

In this paper
we have analyzed the large $N$ limit of the 3-d Gross-Neveu model
on the manifolds $S^2_r \times S^1_\beta$ and $H^2_r \times S^1_\beta$. 
The physical 
observables can be regarded as functions of two parameters, 
the coupling constant of 
the model and the inverse temperature $\beta$. Thus  the critical 
behaviour can be studied with respect to both. In particular,   
we have evaluated the free 
energy, the correlation length and the spontaneous magnetization
  at the UV critical value of the coupling constant, in the 
$\zeta$-function regularization scheme. These quantities  completely 
characterize 
the thermodynamical properties of the model once the 
coupling constant has been fixed. For both manifolds,
the numerical evaluation shows that the
free energy density at the critical coupling is a smooth function of
the temperature; hence no phase transition seems to occur as the 
temperature ($1/\beta$) varies. 

The asymptotic behaviour of the correlation function 
at the critical coupling is also obtained.
We find, in complete analogy with the
results of Ref.\cite{GRV,GV} for the non linear $\sigma$ model, that
the correlation length is finite even for manifolds which are
non compact in all directions ($H^2 \times R$). More precisely,
the {\it finite size} effect is due to the non vanishing curvature
of the manifold which introduces a length scale in the theory.
Finally, we observe two major differences with respect to the non linear 
$\sigma$ model. First, the GN model appears not to be conformally invariant at 
criticality (in the large N limit). Then, we do not observe curvature induced 
symmetry breaking, that is, $m_c$ and $b_c$ stay zero even on curved 
background.

\section*{Acknowledgements}

The authors would like to thank  Giampiero Esposito and Dmitri V. Fursaev
for valuable comments.

\appendix
\section{The algebraic method}
\setcounter{equation}{0}
\noindent
{\it - The space $S^2_r \times S^1_\beta$ -}

\bigskip

\noindent
To solve Eq. \eqn{VI.4} we apply the method described in Ref.\cite{a2}.
According to this method we can construct the eigenvector of the
Dirac operators by means of the eigenvectors, $\phi_{\omega}$,
for the scalar case
\begin{equation}
\Delta_{0} \phi_{\omega} = - \nabla_{\mu}\partial^{\mu} \phi_\omega
= \omega \phi_{\omega}~~~,
\label{a.1}
\end{equation}
and the so-called  {\it covariantly 
constant} spinors, $\epsilon^{\pm}_i$, which are defined in this space 
by the following differential equations
\beqa
\left(\nabla_{\bar{\mu}} \pm {i \over 2 r} \gamma_{\bar{\mu}}\right)
\epsilon^\pm_i &=&0~~~ ~~~\mbox{with}~~~\bar{\mu}=1,2
\label{a.2}\\
\nabla_3 \epsilon^\pm_i &=&0~~~.
\label{a.3}
\eeqa
Note that, the above requirements on the spinors $\epsilon^\pm_i$ are 
compatible with the Bianchi identity which reads
\be
\left[\nabla_{\mu},\nabla_{\nu}\right] \epsilon^{\pm}_i = {1 \over 4}
R_{\mu \nu \lambda \rho} \gamma^{\lambda} \gamma^{\rho} \epsilon^{\pm}_i 
~~~.
\label{a.4}
\ee
The index $i$ just labels the independent solutions of Eqs. 
\eqn{a.2}, \eqn{a.3}. As far as the equation \eqn{a.3} 
is concerned, it only says that the spinors $\epsilon^{\pm}_i$
do not depend on $\theta$. Then we only have to solve the equations
\eqn{a.2} on $S^2_{r}$. The solutions of this problem have been
obtained in Ref.\cite{FM} and are
\begin{equation}
\epsilon^{\pm}_1(\tau,\chi) =\exp\left\{{i\tau \over 2}\right\}~
\left[ \begin{array} {c}
\sin(\chi/2+\pi/4)
\\ \\
\mp\cos(\chi/2+\pi/4)
\end{array}
\right]~~~,
\label{a.5}
\end{equation}
\begin{equation}
\epsilon^{\pm}_2(\tau,\chi) =\exp\left\{{-i\tau \over 2}\right\}~
\left[ \begin{array} {c}
\cos(\chi/2+\pi/4)
\\ \\
\pm\sin(\chi/2+\pi/4)
\end{array}
\right]~~~.
\label{a.6}
\end{equation}
They satisfy the normalization condition
\begin{equation}
\left( \epsilon^{\pm}_{i}\right)^{\dag} \epsilon^{\pm}_{j} = 
\delta_{i j}~~~,
\label{a.7}
\end{equation}
with $i,j=1,2$. It is worth pointing out the right periodicity property 
of $\epsilon^{\pm}_{1(2)}$ spinors
\begin{equation}
\epsilon^{\pm}_i(\tau+2 \pi,\chi)= - \epsilon^{\pm}_i(\tau,\chi)~~~.
\label{a.8}
\end{equation}
However, the same does not occur for the variable $\theta$. To solve this
problem in the reconstruction method we use, for the scalars, twisted
boundary conditions along $\theta$ 
\beqa
\phi_{\omega}(\tau+2\pi,\chi,\theta) & = & \phi_{\omega}(\tau,\chi,\theta)
~~~,\label{a.9}\\
\phi_{\omega}(\tau,\chi,\theta+2 \pi ) & = & - \phi_{\omega}(\tau,\chi,\theta)
~~~.\label{a.10}
\eeqa
The eigenfunctions of (\ref{a.1}) with the above constraints are  then
\be
\phi_{lmn}(\tau,\chi,\theta) = N_{lm} 
\exp\left\{i \left(n+{1 \over 2}\right)\theta\right\}~Y^l_m(\pi/2-\chi,\tau)~~~,
\label{a.11}
\ee
where $Y^l_m$ denote the spherical harmonics, $n\in Z$, $l=0,1,..$,
and $-l \leq m \leq l$ and $N_{lm}$ is a normalization constant. 
The corresponding eigenvalues are 
\be
\omega_{ln} = \omega_l+\omega_n =
{1 \over r^2} l(l+1) + { 1 \over \beta^2} \left(n+{1 \over 2}\right)^2~~~.
\label{a.12}
\ee
In terms of the expressions \eqn{a.5}, \eqn{a.6} and \eqn{a.11}
we can define four independent spinors on $S^2_r \times S^1_\beta$,
for fixed $j$, namely,
\be
|1\rangle \equiv \phi_{lmn} \epsilon_j^{+}~,~~ |2\rangle \equiv
i \gamma^{\bar\mu} (\del_{\bar\mu} \phi_{lmn})
\epsilon_j^{+}~,~~
|3\rangle \equiv \phi_{lmn} \epsilon_j^{-}~,~~|4\rangle \equiv
i \gamma^{\bar\mu} (\del_{\bar\mu} \phi_{lmn}) \epsilon_j^{-}~, 
\label{a.13}
\ee
where $\bar{\mu}=1,2$. It can be can be shown that the action of Dirac
operator is closed on the space spanned by these four vectors, so that
we can find combinations of them which are eigenvectors of the Dirac
operator. In fact, the Dirac operator can be represented as a $4\times
4$ matrix  acting on the subspace spanned by the four vectors 
reported in Eq. \eqn{a.13}
\be
-i \left(\begin{array}{cccc} 
r^{-1}&1&-\frac{2n+1}{2\beta^2}&0\\
{l(l+1) \over r^2} &  0&0&\frac{2n+1}{2\beta} \\
-\frac{2n+1}{2\beta^2}&0&-r^{-1}&1\\
0&\frac{2n+1}{2\beta}&{l(l+1) \over r^2}&0
\end{array} \right)~~~.
\label{a.14}
\ee
The eigenvalues $i \lambda$ are then
\be
\lambda^2 =\left(n+{1 \over 2}\right)^2
{1 \over \beta^2} + {(l+1)^2 \over r^2 }~~~,
~~~\left(n+{1 \over 2}\right)^2
{1 \over \beta^2} + {l^2 \over r^2 }~~~.\label{a.15}
\ee
Note that there are no zero modes. Hence, the eigenvalues of the Dirac operator,
$i \lambda_{ln}^{\pm}$, have the form
\begin{equation}
\lambda_{ln}^{\pm}=\pm \sqrt{
\left(n +{1 \over 2}\right)^2
{1 \over \beta^2} + {(l+1)^2 \over r^2 }}~~~\mbox{with}~~
n\in Z, ~~l=0,1,..
\label{a.16}
\end{equation}
It can be checked, writing the eigenvectors as linear combination of the basis
vectors \eqn{a.13}, that the degeneracy   
is $2(l+1)$.

\bigskip

\noindent
{\it - The space $H^2_r \times S^1_\beta$ -}
\bigskip

\noindent
As for the space $S^2_r \times S^1_\beta$, also in the case of
$H^2_{r} \times S^1_{\beta}$ we look for the solutions of the equations
\beqa
\left(\nabla_{\bar{\mu}} \pm {1 \over 2 r} \gamma_{\bar{\mu}}\right)
\epsilon^\pm_i &=&0~~~ ~~~\mbox{with}~~~\bar{\mu}=1,2
\label{a.17}\\
\nabla_3~ \epsilon^\pm_i &=&0~~~.
\label{a.18}
\eeqa
We find it is easier to solve the above problem in coordinates $\chi\in R$
and $\tau \in [0,2 \pi]$ which are connected to $x$ and $y$ of \eqn{VII.2}
by
\be 
x= {\sinh \chi~ \sin \tau \over \cosh \chi + \sinh \chi ~\cos \tau}~,~~~~
y={1 \over \cosh \chi + \sinh \chi ~\cos \tau}~.
\label{a.19}
\ee
The independent solutions in this case result to be
\begin{equation}
\epsilon^{\pm}_1(\tau,\chi) =\exp\left\{{i\tau \over 2}\right\}~
\left[ \begin{array} {c}
\sinh(\chi/2)
\\ \\
\mp i\cosh(\chi/2)
\end{array}
\right]~~~,
\label{a.20}
\end{equation}
\begin{equation}
\epsilon^{\pm}_2(\tau,\chi) =\exp\left\{{-i\tau \over 2}\right\}~
\left[ \begin{array} {c}
\cosh(\chi/2)
\\ \\
\mp i\sinh(\chi/2)
\end{array}
\right]~~~.
\label{a.21}
\end{equation}
The scalar problem yields to a continuous spectrum \cite{voros,lang}
\be
\omega_{kn}= \omega_{k} + \omega_{n} = {1 \over r^2}
\left( {1 \over 4} + k^2 \right) + { 1 \over \beta^2}
 \left(n +{1 \over 2}\right)^2~~~,
\label{a.22}
\ee
with eigenfunctions 
\be
\phi_{l,m,n} = P_l^m (\cosh\chi) 
\exp\left\{i \left(n+{1 \over 2}\right) \theta + i m \tau \right\}~~~,
\label{a.23}
\ee
and degeneracy \cite{GRV}
\be
\mu (k)= {1\over \pi} \Theta\left(\omega_k-{1\over 4 r^2 }\right) 
\tanh\left(\pi\sqrt{\omega_k-{1\over 4 r^2 }}\right)~~~.
\label{a.24}
\ee
$P_l^m (\cosh \chi)$ are the associated Legendre functions, with 
$ l= -1/2+ik$.

As for $S^2_{r}\times S^1_{\beta}$, also in this case one can construct
the four spinors of Eq. \eqn{a.13} and the action of 
the Dirac operator is closed on the corresponding subspace spanned by them. 
Hence,
the eigenvectors of the Dirac operator can be expressed as a 
linear combination of them. We could in principle proceed in the same way 
as for $S^2_r\times S^1_\beta$ but we choose a simpler approach, using some 
known results. 
It can be easily seen that all the four independent spinors of Eq. \eqn{a.13}
take the form 
\be
|i\rangle \sim |\phi_n\rangle \otimes |\chi_k\rangle~~~, \label{a.25}
\ee
where $  |\phi_n\rangle = (2 \pi \beta)^{-1/2}
\exp \left[ i\left(n+{1\over 2}\right)\theta\right]$ are
eigenfunctions of the scalar Laplacian on $S^1_\beta$ with twisted boundary 
conditions, while $|\chi_k\rangle$ are spinors depending on $H^2_r$ 
coordinates only. Since the eigenvectors of the Dirac operator are linear 
combination of the four spinors \eqn{a.13}, we can also say that 
each eigenvector is of the form \eqn{a.25}.
The action of the squared Dirac operator
\be
\Delta_{1/2}(H_r^2 \times S^1_{\beta})
=\Delta_{1/2}(H^2_r) + \Delta_0(S^1_\beta)
\label{a.26}
\ee
is then factorized in the sense that  
\be
\Delta_{1/2} |\psi_{n,k}\rangle =\Delta_{1/2} |\chi_k\rangle + 
\Delta_{0}|\phi_n\rangle~~~,
\label{a.27} 
\ee
and the heat kernel is just the product of the two heat kernels on $H^2_r$ 
and $S^1_\beta$ (cfr. Ref.\cite{GV}). This is all what we need to write 
down the $\zeta$-function.

\section{Poisson sum formula for $S^2$}
\setcounter{equation}{0}
The general expression for the Poisson sum formula (see Ref.\cite{whitt}) is
\begin{eqnarray}
\sum_{l= -\infty}^{\infty}
\exp\left\{-{4 \pi^2 t \over \gamma^2} (l+\eta)^2
+2 \pi i{x \over \gamma}(l+\eta)\right\} \nn\\
= {\gamma  \over {\sqrt {4 \pi t}}} \sum_{l=-\infty}^{\infty} 
\exp\left\{-{(x+l \gamma)^2 \over 4 t}
-2 \pi i l \eta\right\}
~~.\label{b.1}
\end{eqnarray}
We put $\eta=0$ and $\gamma= 2 \pi$ and derive both the sides of 
\eqn{b.1} with respect to $x$; we get
\be
{i \over 2 \pi} \sum_{l=-\infty}^{\infty} l
\exp\left\{-l^2t + i l x\right\}
=-{1 \over 2 t {\sqrt {4 \pi t}}} \sum_{l=-\infty}^{\infty}  (x+2 \pi l)
\exp\left\{-{(x+ 2 \pi l)^2 \over 4 t}\right\}~.  
\label{b.2}
\ee
Then we observe that 
\be
\sum_{m=-\infty}^{\infty} \mbox{sign}(m) e^{imx}
=1 + 2 i \sum_{m=1}^{\infty} \sin 
(mx) = 1 + i \cot\left({x \over 2}\right), 
\label{b.3}
\ee
so that
\be
\mbox{sign}(m)
={i\over 2 \pi} \int_0^{2 \pi} dx~\cot\left({x \over 2}\right)~
e^{-imx}~.
\label{b.4}
\ee

Now we multiply \eqn{b.2} by ${i \over 2\pi} \cot(x/ 2)$ and integrate
$$
{1 \over 4 \pi^2} \int_0^{2\pi} dx~\sum_{l=-\infty}^{\infty} l 
\cot\left({x\over 2}\right) \exp\left\{-l^2t+ilx\right\}
$$
\be
={i \over (4 \pi t)^{3 \over 2}} 
\int\limits_0^{2 \pi} dx~\cot\left({x \over 2}\right)
\times \sum_{l=-\infty}^\infty (x+2 \pi l) 
\exp\left\{-{(x+2 \pi l)^2 \over 4 t}
\right\}~~~.
\label{b.5}
\ee
Then we use \eqn{b.4} and we get
$$
{1 \over 2 \pi} \sum_{l=-\infty}^{\infty} l~ \mbox{sign}(-l)
\exp\left\{-l^2 t\right\}
$$
\be
=-{1 \over (4 \pi t)^{3 \over 2}} 
\int\limits_0^{2 \pi} dx 
\cot\left({x \over 2}\right) \sum_{l=-\infty}^\infty (x+2 \pi l)
\exp\left\{-{(x+2 \pi l)^2 \over 4 t}\right\}~~~. \label{b.6}
\ee
Performing the change of variable $x'=x+2\pi l$ and exchanging the sum 
with the integral we have
\beqa
\int_0^{2\pi} dx \cot\left({x\over 2}\right) \sum_{l=-\infty}^{\infty} 
(x+ 2\pi l) \exp\left\{-{(x+ 2\pi l)^2\over 4t}\right\}\nn\\
= 
\int_{-\infty}^{\infty} dx 
~x~\cot\left({x \over 2}\right) \exp\left\{-{x^2 \over 4 t}\right\}~.
\label{b.7}
\eeqa
Substituting this expression in \eqn{b.6} we finally get the Poisson 
sum formula for the spin $1/2$ heat kernel on $S^2$
\be
{1\over \pi}\sum_{l=1}^\infty l~ \exp\{-l^2 t\}
={1 \over (4 \pi t)^{3 \over 2}} 
\int_{-\infty}^{\infty} dx ~
x~ \cot\left({x \over 2}\right) \exp\left\{-{x^2 \over 4t}\right\}~~~.  
\label{b.8}
\ee

\newpage

\begin{figure}[]
\epsfysize=9cm
\epsfxsize=14cm
\epsffile{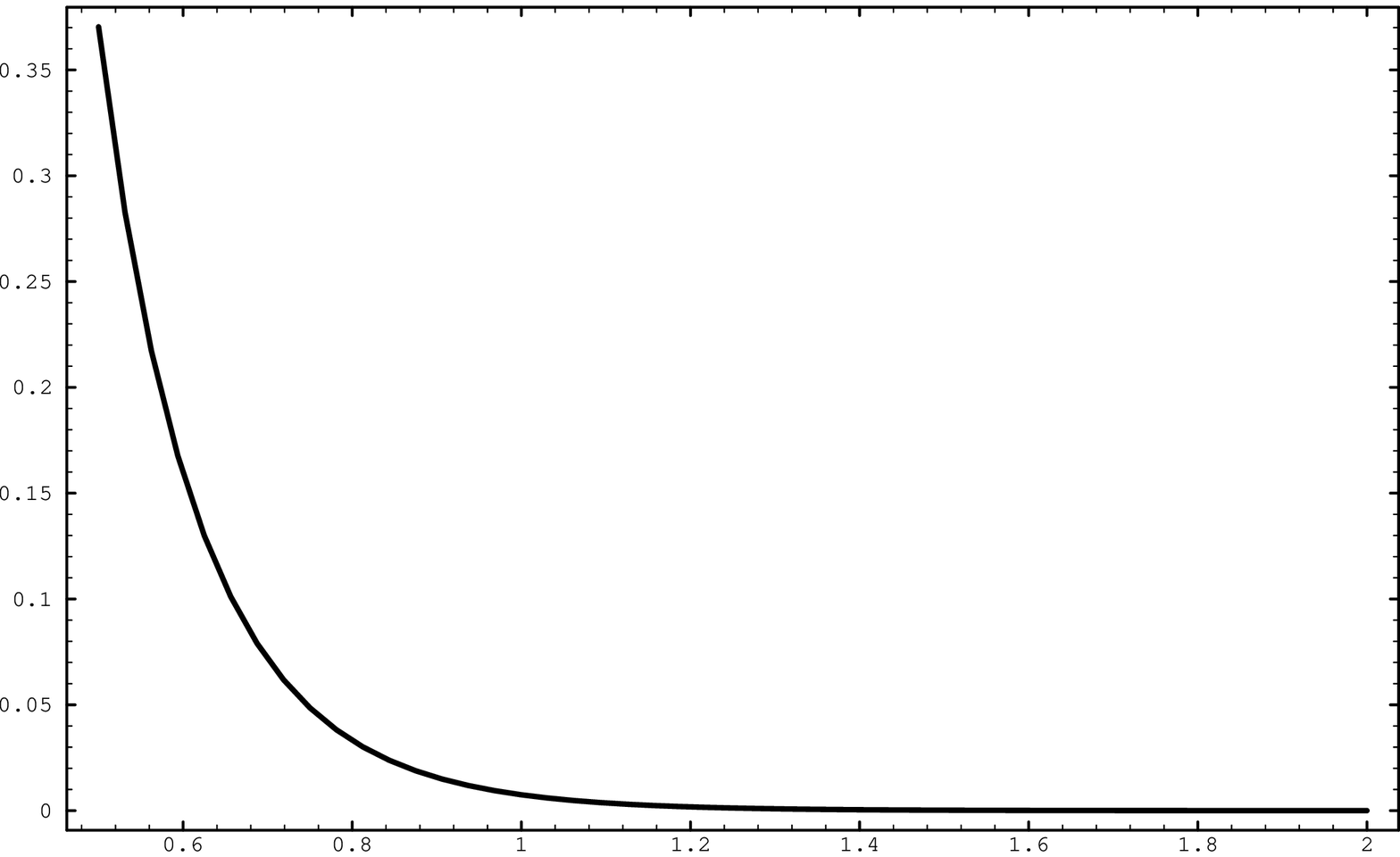}
\caption[]{$S^2_r\times S_\beta^1$. 
The free energy density, $w_c$, of \eqn{VI.19} is plotted
as a function of $\beta/r$, up to the factor
$1/(16 \pi^2 r^3)$.}
\end{figure} 

\begin{figure}[]
\epsfysize=9cm
\epsfxsize=14cm
\epsffile{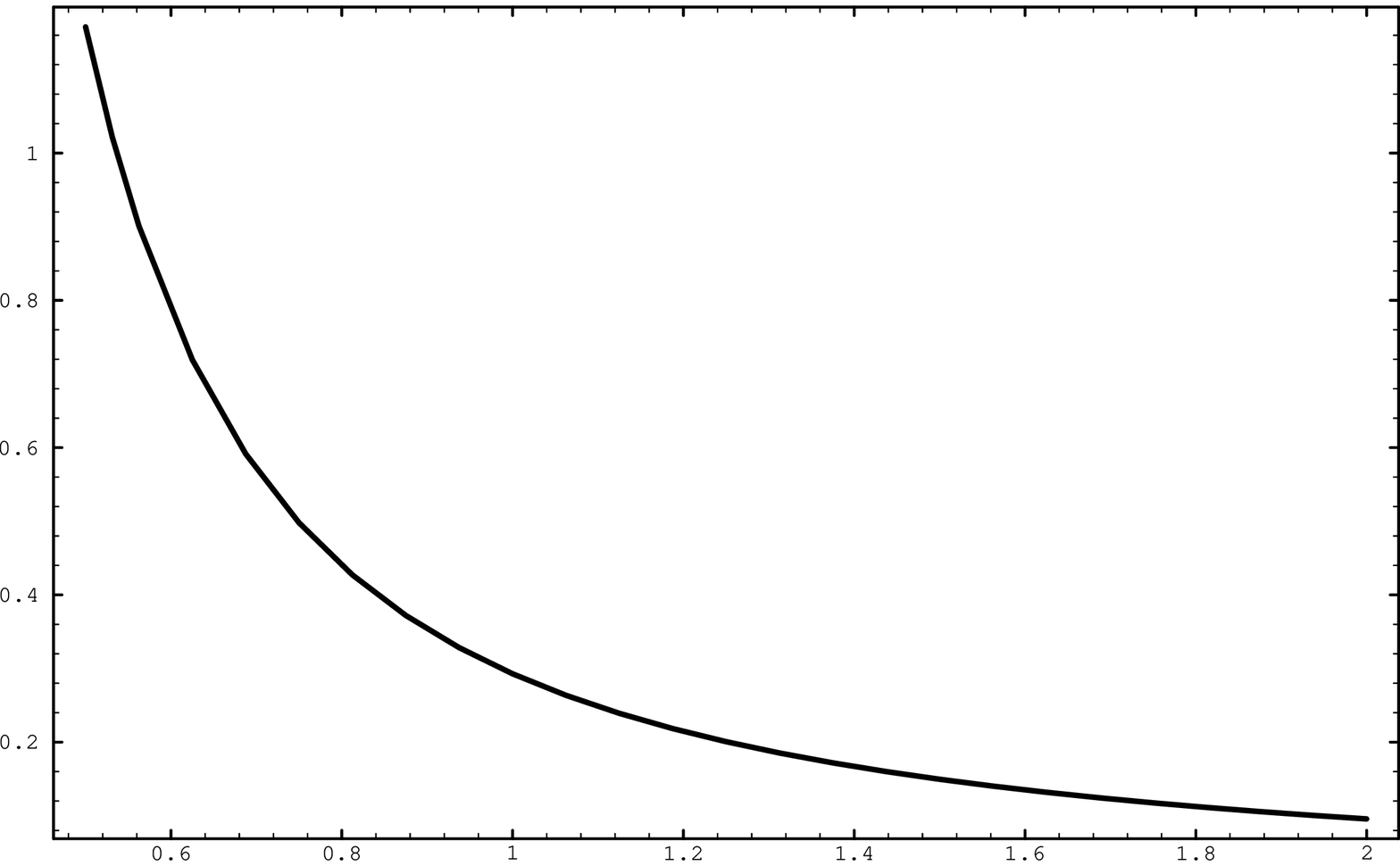}
\caption[]{$H^2_r\times S^1_\beta$. 
The free energy density, $w_c$, of \eqn{VII.22} is plotted
as a function of $\beta/r $, up to the factor
$1/(16 \pi^2 r^3)$.}
\end{figure} 

\begin{thebibliography}{99}

\bibitem{GN}     D. J. Gross and A. Neveu, 
                 \journal{Phys. Rev.}{D10}{74}{3235}.
\bibitem{GRV}    S. Guruswamy, S. G. Rajeev and P. Vitale, \journal{Nucl. 
                 Phys.}{B438}{95}{491}.
\bibitem{GV}     S. Guruswamy and P. Vitale, \journal{Mod. Phys. Let.}   
                 {A11}{96}{1047}.
\bibitem{zinn1}  J. Zinn-Justin, {\it Quantum Field Theory and Critical 
                 Phenomena} (Oxford U. P., Oxford 1989).
\bibitem{jacobs} L. Jacobs, \journal{Phys. Rev} {D10}{74}{3956}
\bibitem{grace1} J. A. Gracey, \journal{Phys. Rev.}{D50}{94}{2840}.
\bibitem{grace2} J. A. Gracey, \journal{Int. J. Mod. Phys.}{A9}{94}{567}.
\bibitem{grace3} J. A. Gracey, \journal{J. Phys.}{A23}{90}{L467}.
\bibitem{zinn2}  J. Zinn--Justin, \journal{Nucl. Phys.}{B367}{91}{105}.
\bibitem{kivel}  N. A. Kivel, A. S. Stepanenko and A. N. Vasilev, 
                 \journal{Nucl. Phys.}{B424}{94}{619}.
\bibitem{VKM}    A.S. Vshivtseev, K. G. Klimenko and V. V. Magnitskii,  
                 \journal{Theor. Math. Phys.}{101}{94}{1436}.
\bibitem{klime}  K. G. Klimenko, \journal{Theor. Math. Phys.}{89}{92}{1161}.
\bibitem{krive}  I. V. Krive and S. A. Naftulin, \journal{Phys. 
                 Rev}{D46}{92}{2737}.
\bibitem{rosen1} B. Rosenstein, B. J. Warr and S. H. Park, \journal{Phys. 
                 Rev. Let.}{62}{89}{1433}.
\bibitem{rosen2} B. Rosenstein, B. J. Warr and S. H. Park, \journal{Phys. 
                  Rev.}{D39}{89}{3088}.
\bibitem{gat}    G. Gat, A. Kovner, B. Rosenstein and B. J. Warr, 
                 \journal{Phys. Let.}{B240}{90}{158}.
\bibitem{kiri}   I. L. Buchbinder and E. N. Kirillova, 
                 \journal{Int. J. Mod. Phys.}{A4}{89}{143}.
\bibitem{kane}   S. Kanemura and H. T. Sato, 
                 \journal{Mod. Phys. Let.}{A11}{96}{785}.
\bibitem{birrel} N.D. Birrell and P.C.W. Davies, {\it Quantum fields in 
                 curved space} (Cambridge U. P., Cambridge 1982).
\bibitem{parisi} G. Parisi, \journal{Nucl. Phys.}{B100}{75}{368}.
\bibitem{gross}  D. J. Gross, in {\it Methods in Field theory} 1975, Les
                 Houches Lectures, R. Balian and J. Zinn--Justin eds.
                 (North--Holland, Amsterdam, 1976).
\bibitem{parker} T. Parker and S. Rosenberg, \journal{J. Diff. Geom.} 
		 {25}{87}{199}.
\bibitem{whitt}  Whittaker and Watson, {\it Modern Analysis} 
                 (Cambridge 1927).
\bibitem{voros}  N. L. Balazs and A. Voros, \journal{Physics Reports}
                 {143}{86}{109}.
\bibitem{doker}  E. D'Hoker and D. H. Phong, \journal{Comm. Math. Phys.}
		 {104}{86}{537}.
\bibitem{a2}     S.M.Christensen, M.J. Duff, G.W. Gibbons and
                 M. Rocek, {\it One loop effects in supergravity with
                 a cosmological constant}, preprint, March 1981, unpublished.
\bibitem{FM}     D.V. Fursaev and G. Miele, {\it Cones, Spins and Heat Kernels},
		 hep-th/9605153 to appear in {\it Nucl. Phys. B}.
\bibitem{lang}   S. Lang, $SL_2(R)$ (Springer, Berlin, 1985).

\end{thebibliography}
\end{document}